\documentclass[aps,prd,twocolumn,eqsecnum]{revtex4}
\begin{document}
\title{Equality of the Hilbert Hamiltonian and the canonical Hamiltonian for gauge theories in a static spacetime}
\author{H. Arthur Weldon}
\email[]{hweldon@WVU.edu}
\affiliation{Department of Physics and Astronomy~West Virginia University,\linebreak
Morgantown, West Virginia, 26506-6315, USA}
\date{\today}

\begin{abstract} The Hilbert energy-momentum tensor for gauge-fixed non-Abelian gauge theories, defined by the variational derivative of the action with respect to the
space-time metric, is a tensor under general  coordinate transformations, symmetric in its indices, and   BRST invariant. The canonical energy-momentum tensor
 has none of these properties but the canonical Hamiltonian does correctly generate the time dependence of the fields.
 It is shown that the Hilbert Hamiltonian $\int d^{3}x\,\sqrt{g}\;T^{0}_{\;\; 0}$ is equal to the canonical Hamiltonian for 
 a general gauge theory coupled to spin 1/2 and spin 0 matter fields (including an $R\phi^{2}$ term) in a static
 background metric ($\partial_{0}g_{\mu\nu}=0$ and $g_{0j}=0$). The equality depends on on the Gauss's law constraint
 but not on the dynamical Euler-Lagrange equations. 
 \end{abstract}

\maketitle

\section{Introduction\label{Intro}}

The Hilbert energy-momentum tensor \cite{Hilbert}  in an arbitrary background metric is determined from the Lagrangian density for matter  and radiation by
\begin{equation}
{\delta \over\delta g^{\mu\nu}(x)}\int d^{4}z\;{\cal L}={\sqrt{g}\over 2}T_{\mu\nu}(x)\label{definition}\end{equation}
where $g\equiv -{\rm det}(g_{\alpha\beta})>0$.
${\cal L}$   transforms as a scalar density under general coordinate transformations (i.e. ${\cal L}/\sqrt{g}$ is a coordinate scalar) and is 
 gauge invariant except for a gauge-fixing term.   $T_{\mu\nu}$  transforms as a tensor
under general coordinate transformations, is symmetric in $\mu\nu$,  
 and is BRST invariant \cite{Becchi,Tyutin}.   It is the source of the gravitational field in   the Einstein field equations. 
 The covariant divergence of the mixed tensor is
\begin{equation}
(T^{\mu}_{\;\;\nu})_{;\mu}={1\over\sqrt{g}}\partial_{\mu}(\sqrt{g}\,T^{\mu}_{\;\;\nu})-{1\over 2}(\partial_{\nu}g_{\alpha\beta})T^{\alpha\beta}.\label{divergence}
\end{equation}
If the fields that appear in ${\cal L}$ are required to satisfy the field equations then $(T^{\mu}_{\;\;\nu})_{;\mu}=0$ (see Sec  94 of \cite{Landau}
or Sec 12.3 of \cite{Weinberg1});
 but it is not    a true conservation law because of the second term in (\ref{divergence}).

The so-called canonical energy-momentum tensor that results from  Noether's first theorem \cite{Noether}  applied to the invariance of ${\cal L}$ under
global spacetime translations is 
\begin{equation}
\sqrt{g}\; \Theta^{\mu}_{\;\;\nu}=\sum_{s}{\partial{\cal L}\over\partial(\partial_{\mu}\chi_{s})}\partial_{\nu}\chi_{s}-\delta^{\mu}_{\;\nu}{\cal L}
\label{canonicalmunu}
\end{equation}
where $\chi_{s}$ runs over all  the fields: gauge bosons, ghosts, spin 1/2 fermions, and scalar bosons. 
 Though $\Theta^{\mu}_{\;\;\nu}$ is not a
coordinate tensor it will be referred to as the canonical energy-momentum tensor because when $g_{\alpha\beta}(x)$ is replaced by the Minkowski metric $(\eta_{\alpha\beta})={\rm diag}(1,-1,-1,-1)$  the resulting quantity is a Lorentz tensor.

For fields obeying Fermi statistics the ordering in (\ref{canonicalmunu}) is not accurate. The ghost fields $\eta_{a}$ and
$\overline{\eta}_{a}$ are independent as are the spin 1/2 fields $\psi$ and $\psi^{\dagger}$; the correct statement of the
first term in (\ref{canonicalmunu}) is
\begin{equation}
{\partial{\cal L}\over\partial(\partial_{\mu}\eta_{a})}\partial_{\nu}\eta_{a}+(\partial_{\nu}\overline{\eta}_{a})
{\partial{\cal L}\over\partial(\partial_{\mu}\overline{\eta}_{a})}
\end{equation}
for ghosts and
\begin{equation}
{\partial{\cal L}\over\partial(\partial_{\mu}\psi)}\partial_{\nu}\psi+(\partial_{\nu}\psi^{\dagger}){\partial{\cal L}\over\partial(\partial(\partial_{\mu}\psi^{\dagger})}
\end{equation}
for spin 1/2 fermions. This is rather cumbersome to repeatedly make explicit and so the simpler form (\ref{canonicalmunu}) will 
often be used.

\paragraph{$\Theta^{\mu}_{\;\;\nu}$ improvements in Minkowski spacetime:} 
The differences between  $T^{\mu}_{\;\;\nu}$ and $\Theta^{\mu}_{\;\;\nu}$  when $g_{\alpha\beta}(x)$ is replaced by the Minkowski metric   was resolved for  electrodynamics  by Belinfante and Rosenfeld  \cite{Belinfante,Rosenfeld}.
In modern approaches   \cite{Weinberg2,Munoz,Saravi,Pons,Blaschke,Ilin,Freese}       an improved $\Theta^{\mu}_{\;\;\nu}$ is obtained by using
both global translation invariance and global Lorentz invariance or with Noether's second theorem
using local translation  invariance. This plus the field equations   lead to an improved Lorentz tensor 
that agrees with the Hilbert tensor in Minkowski spacetime:
  \begin{equation}
T^{\mu\nu}=\Theta^{\mu\nu}
+\partial_{\alpha}\Lambda^{\alpha\mu\nu},\label{Lambda1}
\end{equation}
where the super-potential satisfies $\Lambda^{\alpha\mu\nu}=-\Lambda^{\mu\alpha\nu}$.
Two results follow from  the antisymmetry:
\begin{eqnarray}
T^{0\nu}&=&\Theta^{0\nu}\!+\!\partial_{j}\Lambda^{j0\nu}\label{M1}\\
\partial_{\mu}(T^{\mu\nu})  &=&\partial_{\mu}(\Theta^{\mu\nu}).\label{M2}
\end{eqnarray}
The first shows that 
\begin{equation}\int d^{3}x\, T^{0\nu}= \int d^{3}x\; \Theta^{0\nu}\label{M3};\end{equation}
the second shows that the integrals (\ref{M3}) are time-independent. Both require imposing the field equations.

An interesting generalization is presented in Ref. \cite{Ilin}.  If ${\cal L}$ contains second derivatives or higher of the fields then  $\Lambda^{\alpha\mu\nu}$ is not antisymmetric
in the first two indices; nevertheless $\Lambda^{00\nu}$ is a spatial divergence
which adds another term to (\ref{M1}) so that   (\ref{M3}) is still valid; 
and  $\partial_{\mu}\partial_{\alpha}\Lambda^{\alpha\mu\nu}=0$ so (\ref{M2}) holds which makes (\ref{M3}) time independent.

\paragraph{Derivations of $T_{\mu\nu}$ for an arbitrary metric:} The Hilbert energy-momentum tensor can be derived in a more
geometrical manner using 
the spacetime diffeomorphism group \cite{Gotay} or fibre bundles \cite{Forger}.
Though it is natural to expect that a canonical energy-momentum tensor satisfying the requirements of gauge invariance, $\mu\nu$ symmetry, and covariant conservation
would necessarily be equal to the Hilbert tensor,  \cite{Baker} treats an example of spin 2 fields, the linearized Gauss-Bonnet gravity model, in which 
this is not true.

\paragraph{Static metric with field equations:} 
A static metric satisfies both $\partial_{0}g_{\alpha\beta}=0$ and $g_{0j}=0$. The Schwarzschild and Reissner-Nordstr\"om metrics are of this type
and the geometry shares many features of Minkowski spacetime \cite{Sanchez,Fulling,Derezinski}. 
The vanishing of $g_{0j}$ means that Lagrangian terms like $\sqrt{g}\,g^{\mu\nu}(\partial_{\mu}\phi)(\partial_{\nu}\phi)$, and
analogous terms for gauge bosons and spin 1/2 fermions, do not
violate time-reversal invariance;  the field equations allow separation of variables (time vs  three-space); and the propagators
are invariant under global time translation.  
 If the field equations are satisfied and the metric is static then $\partial_{\mu}(\sqrt{g}\,\Theta^{\mu}_{\;\;0})=0$;  thus the canonical Hamiltonian is time-independent
\begin{equation}
{d\over dt} \int d^{3}x\sqrt{g}\,\Theta^{0}_{\;\;0}=0.\label{canonicalH}\end{equation}
Under the same conditions (static metric plus field equations) the Hilbert tensor satisfies $\partial_{\mu}(\sqrt{g}\,T^{\mu}_{\;\;0})=0$ 
\cite{Landau,Weinberg1} and the Hilbert Hamiltonian is time-independent
\begin{equation}
{d\over dt} \int d^{3}x\sqrt{g}\,T^{0}_{\;\;0}=0.\label{HilbertH}\end{equation}
It is  plausible, but not guaranteed,  that the two Hamiltonians are equal; more importantly,  the argument gives no information
about what happens when the field equations are not satisfied, as is the case in the functional integral formulation of field theory.

\paragraph{Outline:} This paper investigates what happens when the metric is static and only the non-Abelian
Gauss's law is imposed but none of the other field equations.  It is assumed throughout that the field
decrease at spatial infinity is sufficiently rapid as to allow spatial integration by parts with no boundary terms.

Section  II introduces the  Lagrangian density ${\cal L}$  for a general non-Abelian gauge theory
containing five parts:  gauge bosons, gauge fixing, ghosts, spin 1/2 fermions, and scalar bosons: 
\begin{equation}{\cal L}=\sum_{n=1}^{5}{\cal L}^{n}.\end{equation}
The spin 1/2 fermions and the scalar bosons are in arbitrary representations of the gauge group.
The scalar bosons
have Yukawa couplings to fermions and a coupling $\xi R\phi_{i}^{2}$ to the Ricci scalar curvature.
The variational derivative of each action  $\int d^{4}x\,{\cal L}^{n}$ gives the Hilbert energy-momentum
tensor $T_{\mu\nu}^{n}$, which is then evaluated for a static metric. The value of $\sqrt{g}\,g^{00}T_{00}^{n}+{\cal L}^{n}$
is computed for a static metric.

Section III  employs Gauss's  law to obtain
\begin{equation}
\int d^{3}x\sum_{n=1}^{5}\sqrt{g}\,g^{00}T_{00}^{n}=\int d^{3}x\Big[\sum_{s}\Pi_{s}\partial_{0}\chi_{s}-\sum_{n=1}^{5}{\cal L}^{n}\Big],
\end{equation}
which shows the equality of the Hilbert and the canonical Hamiltonians:
\begin{equation}
\int d^{3}x\sqrt{g}\,g^{00}T_{00}=\int d^{3}x\sqrt{g}\, g^{00}\Theta_{00}.\label{H=H}\end{equation}
The dynamical Euler-Lagrange equations are not used. The integrals (\ref{H=H}) could be called 
proto-Hamiltonians since they are time-dependent.

At this point the equality of the two Hamiltonians could  be a special feature of non-Abelian gauge theories, particularly
since the curvature appeared  only in the term $\xi R\phi_{i}^{2}$. To see if the Hamiltonian equality is more general
a term of the form
\begin{equation}
\sqrt{g}\,R^{\mu\nu}(\partial_{\mu}\phi)(\partial_{\nu}\phi)\end{equation}
is investigated. Explicit calculation shows that the  Hilbert energy density and the canonical energy density
are very different but the Hamiltonians are equal.  This rather tedious calculation is
contained in Appendix \ref{example}.

Section IV derives the Hamiltonian equality (\ref{H=H}) in a general manner in which the details of the Lagrangian density
are not specified except that it has only first derivatives of the fields.
The result is
\begin{equation}
\partial_{\mu}(\sqrt{g}\, T^{\mu}_{\;\;0})=\partial_{\mu}(\sqrt{g}\,\Theta^{\mu}_{\;\;0})-\partial_{j}N^{j},\end{equation}
using Gauss's law but not the dynamical field equations.
The spatial integral is
\begin{equation}
{d\over dt}\int d^{3}x\sqrt{g}\,T^{0}_{\;\;0}={d\over dt}\int d^{3}x\sqrt{g}\,\Theta^{0}_{\;\;0}.\end{equation}
Since the fields have arbitrary time dependence the integrals must be equal, which confirms (\ref{H=H})
in the more general case.

\section{Explicit results for the Hilbert energy-momentum tensor}

This section will compute $T_{\mu\nu}^{n}$ for $n=1,\dots 5$ for a general time-dependent metric and then catalogue
for a static metric  the combination $\sqrt{g}\, g^{00}T_{00}^{n}+{\cal L}^{n}$ for gauge bosons, ghosts, and spin 1/2 fermions ($n=1,3,4$)
 and the integrated form $\int d^{3}x [\sqrt{g}\, g^{00}T_{00}^{n}+{\cal L}^{n}]$ for the gauge-fixing term and scalar bosons ($n=2,5$).

\subsection{Gauge bosons}
\paragraph{Gauge bosons in a general metric:}

The covariant field strength tensor
\begin{equation} F^{a}_{\mu\nu}=\partial_{\mu}A_{\nu}^{a}-\partial_{\nu}A_{\mu}^{a}
-ef^{abc}A_{\mu}^{b}A_{\nu}^{c}
\end{equation}
is independent of the metric tensor. ($e$ will be used for the gauge coupling since $g$ is reserved for the
absolute value of the determinant of the metric.). To make the metric dependence explicit ${\cal L}^{1}$ is written in terms of the covariant field strengths
\begin{equation}
{\cal L}^{1}=-{\sqrt{g}\over 4}F_{\mu\alpha}^{a}F_{\nu\beta}^{a}g^{\mu\nu}g^{\alpha\beta}\label{A1}.\end{equation}
The variational derivative of the action gives
\begin{equation}
T_{\mu\nu}^{1}=-F_{\mu\alpha}^{a}F_{\nu\beta}^{a}g^{\alpha\beta}+{g_{\mu\nu}\over 4} F_{\kappa\alpha}^{a}F_{\lambda\beta}^{a} g^{\kappa\lambda}g^{\alpha\beta}.\label{A2}
\end{equation}

\paragraph{Gauge bosons with a static metric:}
In this case \begin{equation}
g^{00}T_{00}^{1}=-{1\over 2}F^{0j}_{a}F^{a}_{0j}+{1\over 4}F_{jk}^{a}F_{a}^{jk},
\end{equation}
and therefore
\begin{equation}
\sqrt{g} \,g^{00}\,T_{00}^{1}+{\cal L}^{1}=-\sqrt{g}\,F^{0j}_{a}F_{0j}^{a}.\label{A}
\end{equation}

\subsection{Gauge-fixing}

\paragraph{Gauge-fixing with general   metric:}
The Lagrange density 
\begin{eqnarray}
{\cal L}^{2}&=&{\lambda\over 2}\sqrt{g}\,[(A^{\mu}_{a})_{;\mu}]^{2}\label{B1}\\
(A^{\mu}_{a})_{;\mu}&=&{1\over\sqrt{g}}\partial_{\mu}(\sqrt{g}\,g^{\mu\nu}A_{\nu}^{a})
 \end{eqnarray}
results in BRST invariance \cite{Becchi,Tyutin}.
Variation of  the action with respect to the metric
 gives for the energy-momentum tensor
\begin{equation}
T_{\mu\nu}^{2}=\lambda\Big[-A_{\mu}^{a}\partial_{\nu}W_{a}-A_{\nu}^{a}\partial_{\mu}W_{a}
+g_{\mu\nu}[{1\over 2}(W^{a})^{2}+A_{a}^{\beta}\partial_{\beta}W_{a}]\Big].\label{B2}
\end{equation}
where for conciseness 
\begin{equation}
W_{a}\equiv (A^{\alpha}_{a})_{;\alpha}={1\over\sqrt{g}}\partial_{\alpha}[\sqrt{g}\,g^{\alpha\beta}A_{\beta}^{a}].
\end{equation}

\paragraph{Gauge-fixing with a static metric:}
Equation (\ref{B2}) with $\mu=\nu=0$ leads to 
\begin{eqnarray}&&\hskip-1cm \sqrt{g}g^{00}\,T_{00}^{2}+{\cal L}^{2}\\
&&=\lambda\sqrt{g}\, [-2A^{0}_{a}\; \partial_{0}W_{a} +(W_{a})^{2}+A^{\beta}_{a}\,\partial_{\beta}W_{a}].\nonumber\label{B3}
\end{eqnarray}
In the term $(W_{a})^{2}$ if one factor of $W_{a}$ is expressed in terms of derivatives then a spatial integration
by parts produces
\begin{eqnarray}&&\hskip-0.5cm \int d^{3}x[\sqrt{g}g^{00}\,T_{00}^{2}+{\cal L}^{2}]\label{B4}\\
&&=\lambda\int d^{3}x\sqrt{g}\, [(\partial_{0}A^{0}_{a}) W_{a}-
A^{0}_{a}\,\partial_{0}W_{a}].\nonumber
\end{eqnarray}

\subsection{Ghost fields}

\paragraph{Ghosts with general time-dependent metric:}
The Lagrangian density for the ghost fields is
\begin{equation}
{\cal L}^{3}=\sqrt{g}\,g^{\mu\nu}(\partial_{\mu}\overline{\eta}_{a})(D_{\nu}\eta)_{a}\label{C1}\end{equation}
where $\overline{\eta}_{a}$ and $\eta_{a}$ obey Fermi statistics, transform in the adjoint representation,
 are not conjugates of each other, and 
\begin{equation}
(D_{\nu}\eta)_{a}=\partial_{\nu}\eta_{a}-ef_{abc}A^{b}_{\nu}\eta_{c}.
\end{equation}
Varying the action with respect to the metric gives
\begin{eqnarray}
T_{\mu\nu}^{3}&=&(\partial_{\mu}\overline{\eta}_{a})(D_{\nu}\eta)_{a}+(\partial_{\nu}\overline{\eta}_{a})(D_{\mu}\eta)_{a}\nonumber\\
&&-g_{\mu\nu}\,g^{\alpha\beta}(\partial_{\alpha}\overline{\eta}_{a})(D_{\beta}\eta)_{a}.\label{C3}
\end{eqnarray}

\paragraph{Ghosts with static metric:}
Equation (\ref{C3}) becomes
\begin{equation}
g^{00}\,T_{00}^{3}=g^{00}(\partial_{0}\overline{\eta}_{a})(D_{0}\eta)_{a}-g^{jk}(\partial_{j}\overline{\eta}_{a})(D_{k}\eta)_{a}
\end{equation}
which leads to
\begin{equation}
\sqrt{g}\,g^{00}\,T_{00}^{3}+{\cal L}^{3}=2\sqrt{g}\, g^{00}(\partial_{0}\overline{\eta}_{a})(D_{0}\eta)_{a}.\label{C}
\end{equation}

\subsection{Spin $1/2$ fermions\label{Fermions}}

\paragraph{Fermions with general  metric:}

The Lagrangian density for fermions
\begin{eqnarray}
{\cal L}^{4}&=&\sqrt{g}\Big\{ {i\over 2}\psi^{\dagger}h\gamma^{\mu}\nabla_{\mu}\psi
-{i\over 2}(\nabla_{\mu}\psi)^{\dagger}h\gamma^{\mu}\psi\nonumber\\
&-&\psi^{\dagger}h(m_{f}+Y_{i}\phi_{i})\psi\Big\}\label{D0}
\end{eqnarray}
requires some explanation. First, $m_{f}$ is a Hermitian mass matrix; $Y_{i}$ are Hermitian Yukawa couplings to the
real scalar fields that are discussed in Sec. \ref{Scalarbosons}.
The spacetime dependent Dirac matrices satisfy
\begin{eqnarray}
&&\big\{\gamma^{\mu},\gamma^{\nu}\}=2g^{\mu\nu}I\nonumber\\
&& (\gamma^{\mu})^{\dagger}=h\gamma^{\mu}h^{-1}\end{eqnarray}
where $h^{\dagger}=h$ is the spin metric \cite{AW1}.  
The spacetime independent Dirac matrices satisfy
\begin{equation} \big\{\gamma^{(\alpha)},\gamma^{(\beta)}\big\}=2\eta^{\alpha\beta}I.\end{equation}
The spacetime dependence of the $\gamma^{\mu}$  is carried by vierbeins,
$\gamma^{\mu}=e^{\mu}_{(\alpha)}\gamma^{(\alpha)}$
where $\eta^{\alpha\beta}e^{\mu}_{(\alpha)}e^{\nu}_{(\beta)}=g^{\mu\nu}$
and $g_{\mu\nu}e^{\mu}_{(\alpha)}e^{\nu}_{(\beta)}=\eta_{\alpha\beta}$.
The spin metric $h=\gamma^{(0)}$ is not a function of spacetime nor is the matrix $\gamma^{5}$:
\begin{equation}
\gamma^{5}=-i\sqrt{g}\,\epsilon_{\alpha\beta\mu\nu}\gamma^{\alpha}\gamma^{\beta}\gamma^{\mu}\gamma^{\nu}/4!\,.\end{equation}
Consequently $\nabla_{\mu}\gamma^{5}=0$ and no additional effort is required if  $\psi_{L}$ and $\psi_{R}$ are in different representations of the gauge group.

The covariant derivative of the fermion field is
\begin{equation}
\nabla_{\mu}\psi=\partial_{\mu}\psi+\Gamma_{\mu}\psi-ieA_{\mu}^{a}T^{a}\psi
\end{equation}
where the spin connection is
\begin{equation}
\Gamma_{\mu}={1\over 8}e^{(\alpha)}_{\rho}\Big(\partial_{\mu}e^{\rho}_{(\beta)}+\Gamma^{\rho}_{\mu\lambda}e^{\lambda}_{(\beta)}\Big)
\big[ \gamma_{(\alpha)}, \gamma^{(\beta)}\big].\label{gamma}
\end{equation}
The detailed calculation of $T_{\mu\nu}^{4}$ is presented in Appendix \ref{fermions} with the result
\begin{eqnarray}
T_{\mu\nu}^{4}&=&{i\over 4}\Big(\psi^{\dagger}h\gamma_{\mu}\nabla_{\nu}\psi
+\psi^{\dagger}h\gamma_{\nu}\nabla_{\mu}\psi\Big)\nonumber\\
&-&{i\over 4}\Big((\nabla_{\nu}\psi)^{\dagger}h\gamma_{\mu}\psi
+(\nabla_{\mu}\psi)^{\dagger}h\gamma_{\nu}\psi\Big) \label{D1}\\
&-& g_{\mu\nu}\,{\cal L}^{4}/\sqrt{g}\,.\nonumber
\end{eqnarray}

\paragraph{Fermions with static metric:}

For a static metric (\ref{D1}) immediately gives
\begin{equation}
\sqrt{g}\,g^{00}\,T_{00}^{4}+{\cal L}^{4}=\sqrt{g}{i\over 2}\Big[\psi^{\dagger}h\gamma^{0}\nabla_{0}\psi
-(\nabla_{0}\psi)^{\dagger}h\gamma^{0}\psi\Big].\label{D2}
\end{equation}
For the static metric $e_{0}^{(0)}=\sqrt{g_{00}}$ and $e_{0}^{(j)}=0$; the spin connection $\Gamma_{0}$
simplifies to
\begin{equation} \Gamma_{0}={1\over 8} (\partial_{j}g_{00})[\gamma^{0},\gamma^{j}].\end{equation}
This satisfies $(\Gamma_{0})^{\dagger}=-h\Gamma_{0}h^{-1}$.  The combination that appears in (\ref{D2}) is
\begin{equation}
h\gamma^{0}\Gamma_{0}-\Gamma_{0}^{\dagger}h\gamma^{0}=h\{\gamma^{0},\Gamma_{0}\}=0.
    \end{equation}
Thus $\nabla_{0}\psi$ in (\ref{D2}) can be replaced by
\begin{equation}
D_{0}\psi=\partial_{0}\psi-ieA_{0}^{a}T^{a}\psi\end{equation}
so that
\begin{equation}
\sqrt{g}\,g^{00}\,T_{00}^{4}+{\cal L}^{4}=\sqrt{g}{i\over 2}\Big[\psi^{\dagger}h\gamma^{0}D_{0}\psi
-(D_{0}\psi)^{\dagger}h\gamma^{0}\psi\Big]\label{D3}
\end{equation}

\subsection{Scalar bosons}\label{Scalarbosons}

\paragraph{Scalars with general metric:} For a set of real scalar fields $\phi_{i}$
the Lagrangian density is
\begin{equation}
{\cal L}^{5}=\sqrt{g}\Big[{g^{\mu\nu}\over 2}(D_{\mu}\phi)_{i}(D_{\nu}\phi)_{i}-U(\phi)-{1\over 2}\xi R\phi_{i}^{2}\Big],\label{E1}
\end{equation}
where the gauge covariant derivative is
\begin{equation}
(D_{\mu}\phi)_{i}=\partial_{\mu}\phi_{i}-ieA_{\mu}^{a}(t^{a})_{ij}\phi_{j}
\end{equation}
with $t^{a}$ imaginary and antisymmetric, $U(\phi)$ is a  polynomial in the fields invariant under local gauge transformations,
$R$ is the Ricci scalar, and $\xi$ is an arbitrary parameter.

The presence of the Ricci scalar $R$ causes some complications in varying the action with respect to the metric tensor.
It is convenient to separate the calculation into two parts:
\begin{equation}
\delta\int d^{4}x\,{\cal L}^{5}={1\over 2}\int d^{4}x\sqrt{g}\,\big[\,T_{\mu\nu}^{5A}\delta g^{\mu\nu}
-\xi(\delta R)\phi_{i}^{2}\big],\label{E3}
\end{equation}
where $T^{5A}_{\mu\nu}$ results from varying everything except $R$:
\begin{eqnarray}
T_{\mu\nu}^{5A}&=&(D_{\mu}\phi)_{i}(D_{\nu}\phi)_{i}\label{E5A}\\
&&\hskip-1cm -g_{\mu\nu}\Big[{1\over 2}g^{\alpha\beta}(D_{\alpha}\phi)_{i}(D_{\beta}\phi)_{i}-U(\phi)-{1\over 2}\xi R\phi_{i}^{2}\Big].\nonumber
\end{eqnarray}
The variation of $R$ required in (\ref{E3}) is available in Sec 10.9 of \cite{Weinberg1}:
\begin{equation}
\delta R=R_{\mu\nu}\delta g^{\mu\nu}-(\delta g^{\mu\nu})_{;\mu;\nu}+g_{\mu\nu}(\delta g^{\mu\nu})^{;\rho}_{\;\; ;\rho}.\label{E5}
\end{equation}
The result of the variation is
\begin{equation}
-{1\over 2}\xi\int d^{4}x\sqrt{g}\,(\delta R)\phi_{i}^{2}={1\over 2}\int d^{4} x\sqrt{g}\,T_{\mu\nu}^{5B}\,\delta g^{\mu\nu},\end{equation}
where
\begin{equation}
T_{\mu\nu}^{5B}=\xi\Big[-R_{\mu\nu}\phi_{i}^{2}+(\phi_{i}^{2})_{;\mu ;\nu}-g_{\mu\nu}(\phi_{i}^{2})^{;\rho}_{\;\; ;\rho}\Big].\label{E5B}
\end{equation}
In summary
\begin{equation}
{\delta\over\delta g^{\mu\nu}}\int d^{4}x\,{\cal L}^{5}={1\over 2}\sqrt{g}\Big[T_{\mu\nu}^{5A}+T_{\mu\nu}^{5B}\Big]\label{E6}
\end{equation}

\paragraph*{Comment:}  $T_{\mu\nu}^{5B}$ has the property that its
 covariant divergence  is
\begin{equation}
(g^{\mu\lambda}T^{5B}_{\lambda\nu})_{;\mu}=-{1\over 2}\xi (\partial_{\nu}R)\phi_{i}^{2}\label{E7}
\end{equation}
with no derivatives of the fields.

\paragraph{Scalars with static metric:} 
 Eq. (\ref{E5A})  with $\mu=\nu=0$ and a static metric gives
\begin{equation}
\sqrt{g}\,g^{00}\,T_{00}^{5A}+{\cal L}^{5}=\sqrt{g}\,g^{00}\,(D_{0}\phi)_{i}(D_{0}\phi)_{i}.\label{E4}
\end{equation}
For a static metric (\ref{E7}) is a true conservation law, $\partial_{\mu}[\sqrt{g}\,g^{00} T^{5B}_{00}]=0$
and so
\begin{equation}
{d\over dt}\int d^{3}x\sqrt{g}\, g^{00}\,T_{00}^{5B}=0.\label{E8}\end{equation}
Since the spatial integral is constant and  the fields have arbitrary spacetime dependence, it is natural to expect that
 the integral  is zero. Appendix \ref{scalars} confirms this:
\begin{equation}
\int d^{3}x \sqrt{g}\,g^{00}T_{00}^{5B}=0.\end{equation}
This and (\ref{E4}) may be summarized as
\begin{eqnarray}
&&\int d^{3}x [\sqrt{g}\,g^{00}\,(T_{00}^{5A}+T_{00}^{5B})+{\cal L}^{5}]\label{E9}\\
 &&\hskip2cm = \int d^{3}x \sqrt{g}\,g^{00}\,(D_{0}\phi)_{i}(D_{0}\phi)_{i}\nonumber.
\end{eqnarray}

{\it Comment:} $T_{\mu\nu}^{1},T_{\mu\nu}^{3},T_{\mu\nu}^{4}, T_{\mu\nu}^{5}$ all agree with Table 1 of  \cite{Forger}, which does 
not include gauge fixing. 

\section{Equality of $\int d^{3}x \,\sqrt{g}\,g^{00}\,T_{00}$ and the canonical Hamiltonian }

The results of the previous five subsections will now be combined. All equations assume a static metric
but allow arbitrary spacetime dependence of the fields.

\subsection{Assembly of the results}

The spatial integral of (\ref{A}), (\ref{C}), and (\ref{D3}) when added to the integrated results (\ref{B4}) and (\ref{E9}) give
\begin{equation}
\int d^{3}x[\sqrt{g}\,g^{00}\,T_{00}+{\cal L}]=\int d^{3}x\sqrt{g}\,[\Omega_{1}+\Omega_{2}],
\label{assemble1}\end{equation}
where $\Omega_{1}$ comes from the gauge boson ${\cal L}^{1}$ and $\Omega_{2}$ from the gauge fixing,
ghosts, spin 1/2 fermions, and scalars:
\begin{eqnarray}
\Omega_{1}&=& -F^{0j}_{a}F^{a}_{0j}\nonumber\\
\Omega_{2}&=&g^{00}\big[\lambda(\partial_{0}A_{0}^{a})W_{a}\!-\!\lambda A_{0}^{a}(\partial_{0}W_{a})+2(\partial_{0}\overline{\eta}_{a})(D_{0}\eta)_{a}\nonumber\\
&+& {i\over 2}\big( \psi^{\dagger}h\gamma_{0}D_{0}\psi-(D_{0}\psi)^{\dagger}h\gamma_{0}\psi\big)\label{G6}\\
&+& (D_{0}\phi)_{i}(D_{0}\phi)_{i}\big]\nonumber
\end{eqnarray}
A more explicit expression of   $\Omega_{1}$ is
\begin{equation}
\sqrt{g}\,\Omega_{1}=\sqrt{g}[-F^{0j}_{a}\partial_{0}A_{j}^{a}+F^{0j}_{a} (D_{j}A_{0})^{a}].\end{equation}
The spatial integral, after an integration by parts, is
\begin{equation}
\int d^{3}x\sqrt{g}\,\Omega_{1}=\!-\!\int d^{3}x [\sqrt{g}\,F^{0j}_{a}\partial_{0}A_{j}^{a}+[D_{j}(\sqrt{g}\,F^{0j})]_{a}A_{0}^{a}].
\end{equation} 
 Gauss' law  requires $[D_{j}(\sqrt{g}\,F^{0j})]_{a}=\sqrt{g}\, J^{0}_{a}$
 where 
 \begin{equation}
 \sqrt{g}\,J^{0}_{a}=-\lambda\sqrt{g}\, g^{00}\partial_{0}W_{a}+\sum_{n=3}^{5}{\partial{\cal L}^{n}\over\partial A_{0}^{a}}\label{J0}
 \end{equation}
and therefore
 \begin{equation}
\int d^{3}x \sqrt{g}\,\Omega_{1}
=-\int d^{3}x \sqrt{g}\,[F^{0j}_{a}\partial_{0}A_{j}^{a}+J^{0}_{a}A_{0}^{a}]\label{O1}.
\end{equation}
 The  charge density  is
\begin{eqnarray}
J^{0}_{a}&=& g^{00}[-\lambda\,\partial_{0}W_{a}+e(\partial_{0}\overline{\eta}_{b})f_{abc}\eta_{c}\\
&+&e\psi^{\dagger}h\gamma_{0}T^{a}\psi-ie(D_{0}\phi)_{i}t^{a}_{ij}\phi_{j}].\nonumber
\end{eqnarray}
Using $J^{0}_{a}$ in (\ref{O1}) and adding this to the spatial integral of $\sqrt{g}\,\Omega_{2}$  produces cancellations of some of the $A_{0}^{a}$ dependence 
and results in
\begin{eqnarray}
&&\hskip-0.5cm \int d^{3}x[\sqrt{g}\,T^{0}_{\;\; 0}+{\cal L}]=\int d^{3}x \sqrt{g}\Big\{\!-\!F^{0j}_{a}\partial_{0}A_{j}^{a}\nonumber\\
&&+g^{00}[\lambda (\partial_{0}A_{0}^{a})W^{a}+(\partial_{0}\overline{\eta}_{a})(\partial_{0}\eta_{a})
+(\partial_{0}\overline{\eta}_{a})(D_{0}\eta)_{a}\nonumber\\
&&+{i\over 2}\big(\psi^{\dagger}h\gamma_{0}\partial_{0}\psi-(\partial_{0}\psi)^{\dagger}h\gamma_{0}\psi\big)\label{cumbersome}\\
&&+(D_{0}\phi)_{i}(\partial_{0}\phi_{i})\Big\}.\nonumber
\end{eqnarray}

\subsection{Canonical momenta}

There are   seven canonical momenta
\begin{eqnarray}
\pi^{j}_{a}&=&{\partial {\cal L}\over \partial(\partial_{0}A_{j}^{a})}=\sqrt{g}\,F^{j0}\nonumber\\
\pi_{a}^{0}&=&{\partial{\cal L}\over\partial(\partial_{0}A_{0}^{a})}=\lambda \sqrt{g}\,g^{00}\,W_{a}\nonumber\\
p_{a}&=&{\partial{\cal L}\over \partial(\partial_{0}\eta_{a})}=\sqrt{g}\,g^{00}\,\partial_{0}\overline{\eta}_{a}\nonumber\\
\overline{p}_{a}&=&{\partial{\cal L}\over \partial(\partial_{0}\overline{\eta}_{a})}=\sqrt{g}\,g^{00}\,(D_{0}\eta)_{a}\label{G3}\\
\pi_{\psi}&=&{\partial{\cal L}\over\partial(\partial_{0}\psi)}=i{\sqrt{g}\over 2}\psi^{\dagger}h\gamma^{0}\nonumber\\
\pi_{\psi^{\dagger}}&=&{\partial{\cal L}\over\partial(\partial_{0}\psi^{\dagger})}=-i{\sqrt{g}\over 2}h\gamma^{0}\psi\nonumber\\
\pi^{\phi}_{i}&=&{\partial{\cal L}\over \partial(\partial_{0}\phi_{i})}=\sqrt{g}\, g^{00}\,(D_{0}\phi)_{i}\,.\nonumber
\end{eqnarray}
The rather cumbersome equation (\ref{cumbersome}) is more recognizable when expressed in terms of the canonical momenta:
\begin{eqnarray}
&&\hskip-0.5cm\int d^{3}x[ \sqrt{g}\,T^{0}_{\;\; 0}+{\cal L}]= \int d^{3}x\Big\{\pi^{j}_{a}\,\partial_{0}A_{j}^{a} +\pi_{a}^{0}\,\partial_{0}A_{0}^{a}\nonumber \\
&&\hskip1.5cm +p_{a}\partial_{0}\eta_{a}+(\partial_{0}\overline{\eta}_{a})\overline{p}_{a}\nonumber \\
&&\hskip1.5cm +\pi_{\psi}\partial_{0}\psi+(\partial_{0}\psi^{\dagger})\pi_{\psi^{\dagger}}+\pi^{\phi}_{i}\partial_{0}\phi_{i}\Big\}.\label{75}
\end{eqnarray}
Of the seven terms involving the canonical momenta the fourth and the sixth have the canonical momenta on the right, as they should be.
Eq. (\ref{75}) may be summarized as
\begin{equation}
\int d^{3}x\sqrt{g}\, T^{0}_{\;\;0}=\int d^{3}x\bigg[\sum_{s}\Pi_{s}\partial_{0}\chi_{s}-{\cal L}\bigg].\label{76}
\end{equation}
The right hand side   is the Legendre transform of the $\partial_{0}\chi_{s}$ dependence of the Lagrangian
to the $\Pi_{s}$ dependence of the canonical Hamiltonian. This proves the equality
\begin{equation}
\int d^{3}x \sqrt{g}\,T^{0}_{\;\; 0}=\int d^{3}x \sqrt{g}\,\Theta^{0}_{\;\;0}\label{equality}\end{equation}
using Gauss's law but not  the dynamical Euler-Lagrange equations.

It is perhaps worth noting that the constraint $[D_{j}(\sqrt{g}\,F^{0j}]_{a}=\sqrt{g}J^{0}_{a}$ may be written in
terms of the canonical momenta
\begin{eqnarray}[D_{j}\pi^{j}]_{a}&=&\pi^{0}_{a}\,\partial_{0}W_{a}+ef_{abc}p_{b}\eta_{c}\nonumber\\
&-&ie\pi_{\psi}T^{a}\psi -ie\,\pi_{i}^{\phi}t^{a}_{ij}\phi_{j}
\end{eqnarray}
and this relation is independent of the metric.

As mentioned in the Introduction, Appendix \ref{example} contains another test of the equality with a Lagrangian density
that contains the Ricci tensor $R^{\mu\nu}$.

\section{General argument}

This section investigates a  more general context in which the form of the Lagrangian density
is not specified, except that it  has  only first derivatives of the fields.
The equality of the two Hamiltonians for a static background metric is again demonstrated.

\paragraph{Divergence of the canonical EMT:} The divergence of the canonical tensor  (\ref{canonicalmunu}) is
\begin{equation}
\partial_{\mu}\big(\sqrt{g}\,\Theta^{\mu}_{\;\;\nu}\big)=\sum_{s}\partial_{\mu}\Big[{\partial{\cal L}\over \partial(\partial_{\mu}\chi_{s})}\partial_{\nu}\chi_{s}\Big]
-\partial_{\nu}{\cal L}.
\end{equation}
The spacetime dependence of ${\cal L}$ occurs in both the fields and the metric:
\begin{equation}
\partial_{\nu}{\cal L}=\sum_{s}\Big[{\partial{\cal L}\over \partial\chi_{s}}(\partial_{\nu}\chi_{s})
+{\partial{\cal L}\over \partial(\partial_{\mu}\chi_{s})}\partial_{\mu}\partial_{\nu}\chi_{s}\Big]-\partial_{\nu}{\cal L}\big|_{\chi}.
\end{equation}
The last term requires differentiating the spacetime dependence of the metric while keeping all the fields $\chi_{s}$ fixed.
Substitution above gives
\begin{eqnarray}
\partial_{\mu}\big(\sqrt{g}\,\Theta^{\mu}_{\;\;\nu}\big)&=&\sum_{s}{\cal M}_{s}(\partial_{\nu}\chi_{s})-\partial_{\nu}{\cal L}\big|_{\chi}\label{candiv}\\
{\cal M}_{s}&\equiv &\partial_{\mu}\Big({\partial{\cal L}\over \partial(\partial_{\mu}\chi_{s})}\Big)-{\partial{\cal L}\over\partial\chi_{s}}.
\end{eqnarray}
${\cal M}_{s}$ vanishes if the Euler-Lagrange equations are imposed but even then $\partial_{\mu}(\sqrt{g}\,\Theta^{\mu}_{\;\;\nu})\neq 0$ 
for a general metric.

\paragraph{Divergence of the Hilbert EMT:} The covariant divergence of $T^{\mu}_{\;\;\nu}$ is computed from the fact that the action is invariant under a coordinate
transformation $x^{\nu}\to x^{\prime\nu}$; see Sec 94 of \cite{Landau} or Sec 12.3 of \cite{Weinberg1}.   The metric and the fields (scalar, ghost, spin 1/2 fermion, and gauge)  transform as follows:
\begin{eqnarray}
g'_{\mu\nu}(x')&=&{\partial x^{\alpha}\over\partial x^{\prime\mu}}{\partial x^{\beta}\over\partial x^{\prime\nu}}\,g_{\alpha\beta}(x)\\
\phi_{i}'(x')&=&\phi_{i}(x)\nonumber\\
\eta_{a}^{\prime}(x')&=&\eta_{a}(x)\nonumber\\
\psi'(x')&=&\psi(x)\\
A_{\alpha}^{\prime a}(x')&=&{\partial x^{\lambda}\over\partial x^{\prime\alpha}}\, A_{\lambda}^{a}(x).\nonumber
\end{eqnarray}
(The covariant vierbein transforms the same as the covariant gauge field but it will not be needed.) Invariance of the action means that
\begin{displaymath}
\int d^{4}x'\,{\cal L}(g_{\mu\nu}'(x'),\chi'(x'))=\int d^{4}x\,{\cal L}(g_{\mu\nu}(x),\chi(x)).
\end{displaymath}
Relabeling of the integration variable gives
\begin{displaymath}
\int d^{4}x\,{\cal L}(g_{\mu\nu}'(x),\chi'(x))=\int d^{4}x\,{\cal L}(g_{\mu\nu}(x),\chi(x)).
\end{displaymath}
The action is invariant under a change in the functional form of the metric and the fields at the same position $x$:
\begin{eqnarray}
\Delta g_{\mu\nu}(x)&\equiv& g'_{\mu\nu}(x)-g_{\mu\nu}(x)\\
\Delta \chi_{s}(x)&\equiv & \chi'_{s}(x)-\chi_{s}(x).
\end{eqnarray}
Let
\begin{equation} x^{\prime\nu}=x^{\nu}+\xi^{\nu}(x)\end{equation}
where $\xi^{\nu}(x)$ is an arbitrary, infinitesimal function. The change in the metric is 
\begin{equation}
\Delta g_{\mu\nu}=-(\xi_{\mu})_{;\nu}-(\xi_{\nu})_{;\mu}.
\end{equation}
For the scalars, ghosts, and spin 1/2 fermions
\begin{equation}
\Delta \chi_{s}=-\xi^{\nu}\partial_{\nu}\chi_{s};\end{equation}
for the vector potential there is an additional term
\begin{equation}
\Delta A_{\alpha}^{a}=-\xi^{\nu}\partial_{\nu}A_{\alpha}^{a}-(\partial_{\alpha}\xi^{\nu})A_{\nu}^{a}.
\end{equation}
Under these variations the action is invariant. The variation with respect to covariant metric gives the contravariant energy-momentum tensor
but with the opposite sign from (\ref{definition}):
\begin{eqnarray}
0\!&=&\!-\!\int d^{4}x{\sqrt{g}\over 2}\,T^{\mu\nu}\Delta g_{\mu\nu}
 +\sum_{s}\int d^{4}x\,{\cal M}_{s}\,\xi^{\nu}\partial_{\nu}\chi_{s}\nonumber\\
 &&\hskip2cm +\int d^{4}x\,{\cal M}^{\mu}_{a}(\partial_{\mu}\xi^{\nu})A_{\nu}^{a}.
 \end{eqnarray}
 In the second term the sum on fields $s$ includes a term for the gauge potential; in the third term the explicit form is
\begin{eqnarray}
{\cal M}_{a}^{\mu}=\partial_{\lambda}\Big({\partial{\cal L}\over\partial(\partial_{\lambda}A_{\mu}^{a})}\Big)-{\partial{\cal L}\over\partial A_{\mu}^{a}}.\label{Mexplicit1}
\end{eqnarray}
An integration by parts in the first and third term yields
\begin{eqnarray}
\int d^{4}x\sqrt{g}\,(T^{\mu\nu})_{:\mu}\xi_{\nu}&=&\sum_{s}\int d^{4}x\,{\cal M}_{s}\,\xi^{\nu}\partial_{\nu}\chi_{s}\nonumber\\
&-&\int d^{4}x \,\partial_{\mu}({\cal M}_{a}^{\mu}A_{\nu}^{a})\xi^{\nu}.
\end{eqnarray}
Since $\xi^{\nu}(x)$ is an arbitrary function, the integrands must be equal:
\begin{equation}
\sqrt{g}\,(T^{\mu}_{\;\;\nu})_{;\mu}=\sum_{s}{\cal M}_{s}(\partial_{\nu}\chi_{s})-\partial_{\mu}({\cal M}_{a}^{\mu}A_{\nu}^{a}).\label{88}
\end{equation}

The first term in (\ref{Mexplicit1}) is $-\partial_{\mu}(\sqrt{g}\, F^{\mu\alpha})_{a}$ which combines with $\partial{\cal L}^{1}/\partial A_{\alpha}^{a}$
to give the gauge covariant derivative $-[D_{\mu}(\sqrt{g}\,F^{\mu\alpha})]_{a}$. The remaining terms from the gauge fixing, ghosts, spin 1/2 fermions,
and scalars define the current density
\begin{equation}
\sqrt{g}\,J^{\alpha}_{a}=-\lambda\partial_{\nu}(\sqrt{g}\,g^{\nu\alpha}W_{a})+ \sum_{n=3}^{5}{\partial{\cal L}^{n}\over\partial A_{\alpha}^{a}}\end{equation}
and so
\begin{equation}
{\cal M}_{a}^{\alpha}=[D_{\mu}(\sqrt{g}\,F^{\mu\alpha})]_{a}+\sqrt{g}\,J^{\alpha}_{a}.\end{equation}
The constraint of Gauss's law requires 
\begin{equation}
{\cal M}^{0}_{a}=0.\label{Gauss}\end{equation}
 Therefore the last term in  (\ref{88}) 
is really only a spatial divergence $\partial_{j}({\cal M}_{a}^{j}A_{\nu}^{a})$ though the form $\partial_{\mu}(M^{\mu}_{a}A_{\nu}^{a})$ will
at times be used  below.

\paragraph{Comparison of divergences of the Hilbert EMT and the canonical EMT.}
The terms  of the form ${\cal M}_{s}(\partial_{\nu}\chi_{s})$ in (\ref{88}) also appear in the canonical divergence (\ref{candiv})
and so (\ref{88}) can be expressed as
\begin{eqnarray}
&&\partial_{\mu}(\sqrt{g}\,T^{\mu}_{\;\;\nu})-{\sqrt{g}\over 2}(\partial_{\nu}g_{\alpha\beta})T^{\alpha\beta}\label{gendiv}\\
&&\hskip1.5cm =\partial_{\mu}(\sqrt{g}\,\Theta^{\mu}_{\;\;\nu})+\partial_{\nu}{\cal L}\big|_{\chi}-\partial_{\mu}({\cal M}_{a}^{\mu}A_{\nu}^{a}).\nonumber
\end{eqnarray}
This holds for a general metric and arbitrary fields using Gauss's law but not the dynamical  Euler-Lagrange equations.

For simplicity suppose ${\cal L}$ has only first and second derivatives of the metric, as is the case for the
general gauge theory in Secs. II and III.
The second and fourth terms of  Eq. (\ref{gendiv}) require the derivatives
\begin{eqnarray}
&&\hskip-0.2cm -{\sqrt{g}\over 2}T^{\alpha\beta}={\partial{\cal L}\over\partial g_{\alpha\beta}}-\partial_{\mu}\bigg[{\partial{\cal L}\over\partial g_{\alpha\beta,\mu}}\bigg]
+\partial_{\mu}\partial_{\rho}\bigg[{\partial{\cal L}\over \partial g_{\alpha\beta,\mu\rho}}\bigg]\nonumber\\
&&\hskip-0.2cm \partial_{\nu}{\cal L}\Big|_{\chi}={\partial{\cal L}\over\partial g_{\alpha\beta}}g_{\alpha\beta,\nu}
+{\partial{\cal L}\over\partial g_{\alpha\beta,\mu}}g_{\alpha\beta,\nu\mu}
+{\partial{\cal L}\over \partial g_{\alpha\beta,\mu\rho}}g_{\alpha\beta,\nu\mu\rho}.\nonumber
\end{eqnarray}
The difference between these two is a total derivative; Eq. (\ref{gendiv}) becomes
\begin{equation}
\partial_{\mu}(\sqrt{g}\,T^{\mu}_{\;\;\nu})
=\partial_{\mu}(\sqrt{g}\,\Theta^{\mu}_{\;\;\nu})+\partial_{\mu}\Sigma^{\mu}_{\;\;\nu}-\partial_{\mu}(A_{\nu}^{a}{\cal M}^{\mu}_{a})\label{divdiv}
\end{equation}
where 
\begin{displaymath}
\Sigma^{\mu}_{\;\;\nu}\equiv {\partial{\cal L}\over \partial g_{\alpha\beta,\mu}}g_{\alpha\beta,\nu}
\!+\!{\partial{\cal L}\over \partial g_{\alpha\beta,\mu\rho}}g_{\alpha\beta,\nu\rho}\!-\! \partial_{\rho}\Big({\partial{\cal L}\over\partial g_{\alpha\beta,\mu\rho}}\Big)g_{\alpha\beta,\nu}.
\end{displaymath}
The derivatives of the metric are not tensors and so $\Sigma^{\mu}_{\;\;\nu}$ is not a tensor.
(If  ${\cal L}$ has any number of metric derivatives of the metric the form (\ref{divdiv}) still holds but $\Sigma^{\mu}_{\;\;\nu}$ is more complicated.)
Because of  Gauss's law (\ref{Gauss}) the spatial integral of (\ref{divdiv}) is
\begin{equation}
{d\over dt} \int d^{3}x\sqrt{g}\,T^{0}_{\;\;\nu}={d\over dt}\int d^{3}x\,\Big[\sqrt{g}\,\Theta^{0}_{\;\;\nu}+\Sigma^{0}_{\;\;\nu}\Big].
\end{equation}
Since the time dependence of the fields and the metric is arbitrary  the integrals must be equal:
\begin{equation}
 \int d^{3}x \sqrt{g}\,T^{0}_{\;\;\nu}=\int d^{3}x \Big[\sqrt{g}\,\Theta^{0}_{\;\;\nu}+\Sigma^{0}_{\;\;\nu}\Big].\label{101}
\end{equation}

\paragraph*{Case 1: Minkowski metric.} If the  metric is chosen to be
$(\eta_{\alpha\beta})={\rm diag}(1,-1,-1,-1)$, then $\Sigma^{\mu}_{\;\;\nu}=0$ in (\ref{101}). The resulting equality
 holds for arbitrary fields and thus is more general than the usual result (\ref{M3}).

\paragraph*{Case 2: Arbitrary time-dependent metric but ${\cal L}$ has no derivatives of the metric.}
These conditions are satisfied by the gauge boson ${\cal L}^{1}$ and by the scalar boson ${\cal L}^{5}$ if the
$\xi R\phi_{i}^{2}$ term is omitted. Obviously $\Sigma^{\mu}_{\;\;\nu}=0$.

\paragraph*{Case 3: Static metric.}
This includes the case of principle interest since ${\cal L}$ for  a general gauge theory contains first derivatives of the metric in the gauge-fixing term
and the spin connection $\Gamma_{\mu}$ for  fermions, and second derivatives of the metric in the $\xi R\phi_{i}^{2}$ term.
For a static metric $\Sigma^{0}_{\;\;0}=0$ and so
\begin{equation}
 \int d^{3}x \sqrt{g}\,T^{0}_{\;\;0}=\int d^{3}x \sqrt{g}\,\Theta^{0}_{\;\;0}.\label{102}
\end{equation}
The dynamical field equations have not been used. 
This agrees with the explicit calculations leading to (\ref{equality}) and explains the miraculous cancellations
for the example considered in Appendix \ref{example}.

\begin{appendix}

\section{Christoffel symbol and curvature tensor for static metric \label{Christoffel}}

A static metric is time-independent and $g_{0j}=0$. Christoffel symbols with an odd number of  time components vanish:
\begin{equation}
\Gamma_{00}^{0}=\Gamma_{k0}^{j}=\Gamma_{k\ell}^{0}=0.\label{Christoffel1}\end{equation}
The Christoffel symbols with two 0's are
\begin{eqnarray}
\Gamma^{j}_{00}&=&-{1\over 2}g^{jk}\partial_{k}g_{00}\label{Christoffel2}\\
\Gamma^{0}_{k0}&=&{1\over 2} g^{00}\partial_{k}g_{00}.\label{Christoffel3}
\end{eqnarray}
$\Gamma^{j}_{k\ell}$ is non-vanishing and independent of $g_{00}$. Two useful contractions are
\begin{eqnarray}
\Gamma^{j}_{jk}&=&{1\over\sqrt{\gamma}}\partial_{k}\sqrt{\gamma}\label{contract1}\\
g^{jk}\Gamma^{\ell}_{jk}&=&-{1\over\sqrt{\gamma}}\partial_{j}(\sqrt{\gamma}\, g^{j\ell})\label{contract2}
\end{eqnarray}
where $|{\rm det}(g_{\alpha\beta})|=g_{00}\,\gamma$.
The Riemann-Christoffel tensor with four spatial components is independent of $g_{00}$:
$^{(4)}R_{ijk\ell}=\, ^{(3)}R_{ijk\ell}$.
If there is one time component
$R_{0jk\ell}=0.$
If there are two time components
\begin{equation}
R_{0j0k}=-{1\over 2}\partial_{j}\partial_{k}g_{00}+{1\over 2}\Gamma^{\ell}_{jk}\partial_{\ell}g_{00}
+{1\over 4} { (\partial_{j}g_{00})(\partial_{k}g_{00})\over g_{00}}.\label{ch2}
\end{equation}
The Ricci tensor with two spatial indices is
\begin{equation}
^{(4)}R_{jk}=g^{00}R_{0j0k}+\,^{(3)}R_{jk}\label{ch3}\end{equation}
where $g^{ab}R_{ajbk}= \,^{(3)}R_{jk}$ is independent of $g_{00}$; 
if there are two time indices
\begin{eqnarray}
^{(4)}R_{00}&=& g^{jk}R_{0j0k}\nonumber\\
&=&-{g_{00}\over 2\sqrt{g}}\partial_{j}\Big[ \sqrt{g}\, g^{jk}{\partial_{k}g_{00}\over g_{00}}\Big];\label{R00}
\end{eqnarray}
and if one time index  $^{(4)}R_{0k}=0$.
The Ricci scalar is
\begin{equation}
^{(4)}R=2g^{00}g^{jk}R_{0j0k}+\, ^{(3)}R\label{ch5}\end{equation}

\section{Calculation of $T_{\mu\nu}^{4}$ for fermions with arbitrary time-dependent metric\label{fermions}}

This appendix contains the detailed calculation of $T_{\mu\nu}^{4}$ displayed in Eq. (\ref{D1}). The fermion field
$\psi$ is not required to satisfy the Dirac equation.

The Lagrangian density (\ref{D0}) may be written
\begin{eqnarray}
{\cal L}^{4}&=&\sqrt{g}{1\over 2}\Big[ \psi^{\dagger}(K\psi)+(K\psi)^{\dagger}\psi\Big]\label{f1}\\
K&=&ih\gamma^{\mu}\nabla_{\mu}-h(m_{f}+Y_{i}\phi_{i})\nonumber.
\end{eqnarray}
The variation of  ${\cal L}^{4}$  with respect to the metric is
\begin{eqnarray}
\delta{\cal L}^{4}&=&- \;{1\over 2} g_{\mu\nu}\,\delta g^{\mu\nu}\,{\cal L}^{4}\label{f2}\\
&+& {1\over 2}\sqrt{g}\Big[\psi^{\dagger}(\delta K\psi)+(\delta K\psi)^{\dagger}\psi\Big]\nonumber.
\end{eqnarray}
In \cite{AW1} the variations are shown to be
\begin{eqnarray}
\delta K&=&{i\over 2}(\delta g^{\mu\nu})h\gamma_{\mu}\nabla_{\nu}+(\delta G)K +K(\delta G)+\delta X\nonumber\\
\delta G&=&{1\over 8} \eta^{\alpha\beta}e_{(\alpha)}^{\lambda}[\gamma_{\lambda},\gamma_{\nu}]\,\delta e_{(\beta)}^{\nu}\label{f3}\\
\delta X&=& {i\over 8} (\delta \Gamma_{\lambda\nu}^{\mu})h\gamma^{\lambda}[\gamma_{\mu},\gamma^{\nu}]\nonumber.
\end{eqnarray}
Because $(\delta X)^{\dagger}=-\delta X$
it disappears from (\ref{f2}) and so
\begin{eqnarray}
\delta {\cal L}^{4}&=&-{1\over 2} g_{\mu\nu}\,{\cal L}^{4}\,\delta g^{\mu\nu}\nonumber\\
&&\hskip-1cm +{i\over 4}\sqrt{g}\Big[\psi^{\dagger}h\gamma_{\mu}\nabla_{\nu}\psi -(\nabla_{\nu}\psi)^{\dagger}h\gamma_{\mu}\psi\Big]
\delta g^{\mu\nu}\nonumber\\
&+&{1\over 2} \Big[\Delta_{1}+\Delta_{2}\Big]\label{f4}\\
\Delta_{1}&\equiv&\sqrt{g}\Big[\psi^{\dagger}(\delta G K\psi)+(\delta G K\psi)^{\dagger}\psi\Big]\nonumber\\
\Delta_{2}&\equiv&\sqrt{g}\Big[\psi^{\dagger}(K\delta G\psi)+(K\delta G\psi)^{\dagger}\psi\Big]\nonumber .
\end{eqnarray}
$\Delta_{2}$ contains derivatives of the vierbein because of $K\delta G$.
To simplify $\Delta_{2}$ the following identity valid for arbitrary spinor fields $\psi_{1}$ and $\psi_{2}$, is useful
\begin{equation}
\sqrt{g} \,\psi_{1}^{\dagger}(K\psi_{2})-\sqrt{g}(K\psi_{1})^{\dagger}\psi_{2}=-\partial_{\lambda}(\sqrt{g}\,\psi_{1}^{\dagger}h\gamma^{\lambda}\psi_{2})
.\label{f5}\end{equation}
For the first term in $\Delta_{2}$ take $\psi_{1}=\psi$ and $\psi_{2}=\delta G\psi$; for the second term in $\Delta_{2}$ take
$\psi_{1}=K\delta G\psi$ and $\psi_{2}=\psi$: the result is 
\begin{equation}
\Delta_{2}=\Delta_{1}+i\partial_{\lambda}\Big[\sqrt{g}\,\psi^{\dagger}h\gamma^{\lambda}\delta G\psi
-\sqrt{g}(\delta G\psi)^{\dagger}h\gamma^{\lambda}\psi\Big].
\end{equation}
Since $(\delta G)^{\dagger}=-h(\delta G)h^{-1}$ this is
\begin{equation}
\Delta_{2}=\Delta_{1}+i\partial_{\lambda}\Big[\sqrt{g}\,\psi^{\dagger}h\{\gamma^{\lambda},\delta G\}\psi\Big].
\end{equation}
The variation of ${\cal L}^{4}$ becomes
\begin{eqnarray}
\delta {\cal L}^{4}&=&-{1\over 2}g_{\mu\nu}\delta g^{\mu\nu}{\cal L}^{4}\nonumber\\
&+&i{\sqrt{g}\over 4}\big[\psi^{\dagger}h\gamma_{\mu}\nabla_{\nu}\psi-(\nabla_{\nu}\psi)^{\dagger}h\gamma_{\mu}\psi
\big]\delta g^{\mu\nu}\nonumber\\
&+& \sqrt{g}\Big[ \psi^{\dagger}(\delta G)K\psi+(K\psi)^{\dagger}(\delta G)^{\dagger}\psi\big]\label{f8}\\
&+&i\partial_{\lambda}\Big[\sqrt{g}\,\psi^{\dagger}h\{\gamma^{\lambda},\delta G\}\psi\Big]\nonumber.
\end{eqnarray}
The total derivative in the fourth line does not contribute to the variation of the action. From (\ref{f3}) the  derivative of $G$ with respect to the vierbein is 
\begin{equation}
e_{\mu(\beta)}{\partial G\over\partial e^{\nu}_{(\beta)}}={1\over 8}[\gamma_{\mu},\gamma_{\nu}]\label{f9},
\end{equation}
and the  derivative  respect to the metric tensor is
\begin{equation}
{\partial G\over\partial g^{\mu\nu}}={1\over 2}\Big[e_{\mu(\beta)}{\partial G\over\partial e^{\nu}_{(\beta)}}
+e_{\nu(\beta)}{\partial G\over\partial e^{\mu}_{(\beta)}}\Big] =0\label{f10}.
\end{equation}
Consequently the energy-momentum tensor is determined by just the first  two lines of (\ref{f8}):
\begin{eqnarray}
T_{\mu\nu}^{4}&=&{i\over 4}\Big[\psi^{\dagger}h\gamma_{\mu}\nabla_{\nu}\psi
+\psi^{\dagger}h\gamma_{\nu}\nabla_{\mu}\psi\Big]\nonumber\\
&-&{i\over 4}\Big[(\nabla_{\nu}\psi)^{\dagger}h\gamma_{\mu}\psi
+(\nabla_{\mu}\psi)^{\dagger}h\gamma_{\nu}\psi\Big]\label{f11}\\
&-&g_{\mu\nu}\,{\cal L}^{4}/\sqrt{g}\nonumber.
\end{eqnarray}

\section{Proof that $\int d^{3}x\sqrt{g}\,g^{00}\,T_{00}^{5B}=0$\label{scalars}}

The ``extra" piece (\ref{E5B})  in the energy-momentum tensor for scalar bosons is\begin{equation}
T_{\mu\nu}^{5B}=-R_{\mu\nu}\Psi+\Psi_{;\mu ;\nu}-g_{\mu\nu}\Psi^{;\rho}_{\;\; ;\rho}\label{Scalar1}.
\end{equation}
where $\Psi=\xi\phi_{i}^{2}$.
This came from the term $-{1\over 2}\xi \sqrt{g}\, (\delta R)\phi_{i}^{2}$ in (\ref{E3}). The following applies to any action  of the form
\begin{equation} -{1\over 2}\int d^{4}x \sqrt{g}\, \Psi\, R\label{s5B2}\end{equation} 
where $\Psi$ is a coordinate scalar. A Lagrangian density of the form $\sqrt{g}\, X^{\mu}\partial_{\mu}R$ or $\sqrt{g}\, Y^{\mu\nu} R_{;\mu;\nu}$
has an action of the form (\ref{s5B2}) after an integration by parts.

\paragraph*{Static metric:}
To analyze $T_{00}^{5B}$ for a static metric,
organize the $\mu=\nu=0$ component  as
\begin{eqnarray}
\sqrt{g}\,g^{00}\,T_{00}^{5B}&=&A^{0}_{\;\;0}+B^{0}_{\;\;0}\nonumber\\
A^{0}_{\;\;0}&=&-{\sqrt{g}\over g_{00}}R_{00}\Psi\\
B^{0}_{\;\;0}&=&\sqrt{g}\big[\Psi^{;0}_{\;\; ;0}-\Psi^{;\lambda}_{\;\; ;\lambda}\big]\nonumber.
\end{eqnarray}
From (\ref{R00})
\begin{equation}
A^{0}_{\;\;0}={1\over 2}\partial_{j}\Big[\sqrt{g}\, g^{jk}{\partial_{k}g_{00}\over g_{00}}\Big]\Psi.\label{APsi}
\end{equation}
In $B^{0}_{\;\;0}$  the time derivatives of the fields all cancel and leave
\begin{equation}
B^{0}_{\;\;0}=-{\sqrt{g}\over g_{00}}\Gamma_{00}^{k}\partial_{k}\Psi-\partial_{k}\big[\sqrt{g}\,g^{kj}\partial_{j}\Psi)\big].
\end{equation}
Using (\ref{Christoffel2}) gives a more complicated looking result:
\begin{equation}B^{0}_{\;\;0}={1\over 2}\sqrt{g}\,g^{jk}{\partial_{j}g_{00}\over g_{00}}\partial_{k}\Psi
-\partial_{k}\big[\sqrt{g}\, g^{jk}\partial_{j}\Psi)\big]\end{equation}
but the sum with $A^{0}_{\;\;0}$ gives another total derivative:
\begin{equation}
A^{0}_{\;\;0}+B^{0}_{\;\;0}={1\over 2}\partial_{j}\Big[\sqrt{g}\, g^{jk}{\partial_{k}g_{00}\over g_{00}}\Psi\Big]-\partial_{k}\big[\sqrt{g}\, g^{jk}\partial_{j}\Psi)\big].
\end{equation}
Consequently
\begin{equation}
\int d^{3}x \sqrt{g}\,g^{00}T_{00}^{5B}=0.\end{equation}

\section{A complicated example \label{example}}

This appendix presents the details of the example mentioned in Sec. I in which ${\cal L}$ depends on the Ricci curvature tensor $R^{\mu\nu}$. Despite the
complications the result is again that the Hilbert Hamiltonian is equal to the canonical Hamiltonian for a static metric even without using the field equation.
The Lagrangian density is
\begin{equation}
{\cal L}=\sqrt{g}\,R^{\mu\nu}(\partial_{\mu}\phi)(\partial_{\nu}\phi)\end{equation}
and could be added to the conventional Lagrangian density for a scalar field if  multiplied by
a coefficient of mass dimension $M^{-2}$. 
The question is not whether this ${\cal L}$ is a physically acceptable addition but whether the resulting Hamiltonians, Hilbert and canonical, are equal.
The canonical energy-momentum pseudo-tensor  (\ref{canonicalmunu}) is
\begin{equation}
\sqrt{g}\,\Theta^{\mu}_{\;\;\nu}=2\sqrt{g}\,R^{\mu\alpha}(\partial_{\alpha}\phi)(\partial_{\nu}\phi)-\delta^{\mu}_{\;\;\nu}{\cal L}.
\end{equation}
For a static background metric 
\begin{equation}
\Theta^{0}_{\;\ 0}=R^{00}\partial_{0}\phi\,\partial_{0}\phi-R^{jk}\partial_{j}\phi\,\partial_{k}\phi.\label{Counter3}
\end{equation}
The field equations have not been used.

\paragraph{Hilbert $T_{\mu\nu}$ for a general metric:}
The Hilbert energy-momentum tensor for a general metric is more complicated to compute. It is convenient to organize ${\cal L}$ as
\begin{equation}
{\cal L}=\sqrt{g}\,R_{\alpha\beta}\,g^{\alpha\mu}g^{\beta\nu}\,\Phi_{\mu\nu},\end{equation}
where $\Phi_{\mu\nu}=(\partial_{\mu}\phi)(\partial_{\nu}\phi)$ is a tensor independent of the metric.
One needs the variation of the covariant Ricci tensor with respect to the contravariant $\delta g^{\mu\nu}$
given in Eq. (10.9.3) of \cite{Weinberg1}: \begin{eqnarray}
\delta R_{\alpha\beta}&=&{1\over 2}g_{\mu\nu}(\delta g^{\mu\nu})_{;\alpha;\beta}
+{1\over 2} g_{\alpha\mu}g_{\beta\nu}(\delta g^{\mu\nu})^{;\lambda}_{\;\;;\lambda}\nonumber\\
&-&{1\over 2} g_{\alpha\mu}(\delta g^{\mu\nu})_{;\beta;\nu}
-{1\over 2} g_{\beta\nu}(\delta g^{\mu\nu})_{;\alpha;\mu}.
\end{eqnarray}
The variational derivative of the action allows the
covariant derivatives of $\delta g^{\mu\nu}$ to be shifted to the the fields and leads to
\begin{eqnarray}
T_{\mu\nu}&=&2\Phi_{\mu\alpha}R^{\alpha}_{\;\;\nu}+2\Phi_{\nu\alpha}R^{\alpha}_{\;\;\mu}-g_{\mu\nu}\Phi_{\alpha\beta}R^{\alpha\beta}\nonumber\\
&+&g_{\mu\nu}(\Phi_{\alpha\beta})^{;\alpha;\beta}+(\Phi_{\mu\nu})^{;\lambda}_{\;\; ;\lambda}\\
&-&(\Phi_{\mu\alpha})_{;\nu}^{\;\; ;\alpha}-(\Phi_{\nu\alpha})_{;\mu}^{\;\; ;\alpha}.\nonumber
\end{eqnarray}
This holds for a general metric.

\paragraph{Hilbert $T_{00}$ for a static metric:}
Specialize to a static metric, set $\mu=\nu=0$, and use $R^{0}_{\;\;j}=R^{j}_{\;\; 0}=0$ to obtain
\begin{eqnarray}
T_{00}&=& P_{00}+Q_{00}\nonumber\\
P_{00}&=&  g_{00}\Big[3R^{00}\Phi_{00}-R^{jk}\Phi_{jk}\Big] \\
Q_{00}&=&g_{00} (\Phi_{\alpha\beta})^{;\alpha;\beta}+(\Phi_{00})^{;\lambda}_{\;\;;\lambda}- 2(\Phi_{0\alpha})_{;0}^{\;\; ;\alpha}.\nonumber
\end{eqnarray}
Note that $P_{00}$ contains a factor $3$ not present in $\Theta_{00}$.
In $Q_{00}$ all the double time derivatives cancel and leave
\begin{displaymath}
Q_{00}=g_{00}[(\Phi_{0\ell})^{;\ell;0}-(\Phi_{0\ell})^{;0;\ell}]+g_{jk}(\Phi_{00})^{;j;k}+g_{00}(\Phi_{jk})^{;j;k}.
\end{displaymath}
It is convenient to change  contravariant derivatives to covariant derivatives
\begin{displaymath}
Q_{00}=g_{00}[(\Phi^{0\ell})_{;\ell;0}-(\Phi^{0\ell})_{;0;\ell}]+g^{jk}(\Phi_{00})_{;j;k}+g_{00}(\Phi^{jk})_{;j;k}.
\end{displaymath}
The term in square brackets may be evaluated in terms of the curvature tensor
\begin{eqnarray}
(\Phi^{0\ell})_{;\ell;0}-(\Phi^{0\ell})_{;0;\ell}&=&-R^{\ell}_{\;\; 0\ell 0}\Phi^{00}+R^{0}_{\;\; j0k}\Phi^{jk}\nonumber\\
&=&-R_{00}\Phi^{00}+[R_{jk}-R^{\ell}_{\;\; j\ell k}]\Phi^{jk}\nonumber
\end{eqnarray}
 and therefore
  \begin{eqnarray}
 P_{00}+Q_{00}&=&g_{00}\,2R^{00}\Phi_{00}
 -g_{00}R^{\ell}_{\;\; j\ell k}\Phi^{jk}\\
 &+&g^{jk}(\Phi_{00})_{;j;k}+g_{00}(\Phi^{jk})_{;j;k}.\nonumber
 \end{eqnarray}
The difference between the Hilbert energy density and the canonical energy density is 
\begin{eqnarray}
\sqrt{g}\,[T^{0}_{\;\; 0}-\Theta^{0}_{\;\; 0}]&=&\sqrt{g}\,[R^{00}\Phi_{00}+R^{jk}\Phi_{jk}] \\
&+&X_{1} +X_{2} +X_{3}\label{PQ1Q2}\nonumber\\ 
X_{1}&\equiv&\sqrt{g}\,g^{jk}(\Phi^{0}_{\;\;0})_{;j;k}\nonumber\\
X_{2}&\equiv& \sqrt{g}\, (\Phi^{jk})_{;j;k}\label{PQQ}\\
X_{3}&\equiv&-\sqrt{g}\, R^{\ell}_{\;\; j\ell k}\Phi^{jk}\nonumber
\end{eqnarray}

\paragraph{Analysis of $X_{1}$:} 
To simplify $X_{1}$  use (\ref{Christoffel1}) and (\ref{contract2}) to obtain
\begin{eqnarray}
g^{jk}(\Phi^{0}_{\;\; 0})_{;j;k}&=&g^{jk}[\partial_{k}(\Phi^{0}_{\;\; 0})_{j}-\Gamma_{jk}^{\ell}(\Phi^{0}_{\;\;  0})_{;\ell}]\nonumber\\
&=&g^{jk}\partial_{k}(\Phi^{0}_{\;\;0})_{;j}+{1\over\sqrt{\gamma}}\partial_{k}(\sqrt{\gamma}\,g^{k\ell})(\Phi^{0}_{\;\;0})_{;\ell}\nonumber\\
&=&{1\over\sqrt{\gamma}}\partial_{k}[\sqrt{\gamma}\,g^{jk}(\Phi^{0}_{\;\;0})_{;j}].
\end{eqnarray}
Therefore
\begin{equation}
X_{1}=\sqrt{g_{00}}\,\partial_{k}[\sqrt{\gamma}\,g^{jk}(\Phi^{0}_{\;\;0})_{;j}].
\end{equation}
In the volume integral of $X_{1}$ a spatial integration by parts gives
\begin{equation} 
\int d^{3}x\,X_{1} = -{1\over 2} \int d^{3}x\,\sqrt{g}\,g^{jk}{\partial_{k}g_{00}\over g_{00}}(\Phi^{0}_{\;\;0})_{;j}.
\end{equation}
Since $(\Phi^{0}_{\;\;0})_{;j}=\partial_{j}\Phi^{0}_{\;\;0}$ another integration by parts gives
\begin{equation}
\int d^{3}x\,X_{1} ={1\over 2} \int d^{3}x\,\partial_{j}\Big[\sqrt{g}\,g^{jk}{\partial_{k}g_{00}\over g_{00}}\Big]\Phi^{0}_{\;\;0}.
\end{equation}
From the identity (\ref{R00}) this is
\begin{eqnarray}
\int d^{3}x\,X_{1}&=&-\int d^{3}x \sqrt{g}\, R^{0}_{\;\;0}\Phi^{0}_{\;\;0}\nonumber\\
&=&-\int d^{3}x \sqrt{g}\, R^{00}\Phi_{00}.\label{P}
\end{eqnarray}

\paragraph{Analysis of $X_{2}$:}
To simplify $X_{2}$ use (\ref{contract1}) 
\begin{eqnarray}
(\Phi^{jk})_{;j;k}&=&\partial_{k}\big[(\Phi^{jk})_{;j}\big]+\Gamma^{k}_{k\ell}(\Phi^{j\ell})_{;j}\nonumber\\
&=&\partial_{k}\big[(\Phi^{jk})_{;j}\big]+{1\over\sqrt{\gamma}}(\partial_{\ell}\sqrt{\gamma})(\Phi^{j\ell})_{;j}\nonumber\\
&=&{1\over\sqrt{\gamma}}\partial_{k}\big[\sqrt{\gamma}\,(\Phi^{jk})_{;j}\big];
\end{eqnarray}
therefore
\begin{equation}
X_{2}=\sqrt{g_{00}}\,\partial_{k}[\sqrt{\gamma}\,(\Phi^{jk})_{;j}].
\end{equation}
In the volume integral of $X_{2}$ a spatial integration by parts yields
\begin{equation}
\int d^{3}x\,X_{2}=-{1\over 2}\int d^{3}x \sqrt{g}\,(\Phi^{jk})_{;j}{\partial_{k}g_{00}\over g_{00}}.\label{Q1}
\end{equation}
The necessary covariant derivative is
\begin{eqnarray}
(\Phi^{jk})_{;j}&=&\partial_{j}\Phi^{jk}+\Gamma_{j\ell}^{j}\Phi^{\ell k}+\Gamma_{j\ell}^{k}\Phi^{j\ell}\nonumber\\
&=&{1\over\sqrt{\gamma}}\partial_{j}\big[\sqrt{\gamma}\,\Phi^{jk}\big]+\Gamma_{j\ell}^{k}\Phi^{j\ell}.
\end{eqnarray}
Substitution into (\ref{Q1}) and integration by parts gives
\begin{eqnarray}
\int d^{3}x\;X_{2}&=&\int d^{3}x{\sqrt{g}\over g_{00}}\Big[{\partial_{j}\partial_{k}g_{00}\over 2}-{\partial_{j}g_{00}\partial_{k}g_{00}\over 4g_{00}}\Big]\Phi^{jk}\nonumber\\
&-&{1\over 2}\int d^{3}x\,{\sqrt{g}\over g_{00}}\,(\partial_{\ell}g_{00})\Gamma^{\ell}_{jk}\Phi^{jk}.
\end{eqnarray}
Comparison with (\ref{ch2}) shows that
\begin{eqnarray}
\int d^{3}x \,X_{2}&=&-\int d^{3}x\sqrt{g}\,g^{00}\,R_{0j0k}\Phi^{jk}.
\end{eqnarray}
Combining this with $X_{3}$ from (\ref{PQQ}) gives
\begin{eqnarray}
\int d^{3}x \,[X_{2}+X_{3}]&=&-\int d^{3}x\sqrt{g}\,[R^{0}_{\;\; j0k}+R^{\ell}_{\;\; j\ell k}]\Phi^{jk}\nonumber\\
&=&-\int d^{3}x\sqrt{g}\,R_{jk}\Phi^{jk}\label{Q1Q2}
\end{eqnarray}

\paragraph{Result:}
From (\ref{P}) and (\ref{Q1Q2}) 
\begin{eqnarray}
\int d^{3}x[\sqrt{g}\,R^{00}\Phi_{00}+X_{1}]&=&0\\
\int d^{3}x\,[\sqrt{g}\,R^{jk}\Phi_{jk}+X_{2}+X_{3}]&=&0.
\end{eqnarray}
And so from (\ref{PQ1Q2}) 
\begin{equation}
\int d^{3}x\sqrt{g}\,[T^{0}_{\;\; 0}-\Theta^{0}_{\;\; 0}]=0\end{equation}
which shows that the Hilbert and canonical Hamiltonians are equal.

\end{appendix}


\begin{thebibliography}{}


\bibitem{Hilbert} D. Hilbert, Die Grundlagen der Physik,   Nachr. Ges. Wiss. Gott., Math-Phys., Kl. {\bf 27}, 395 (1915).

\bibitem{Becchi} C. Becchi, A. Rouet, and R. Stora, Renormalization of the abelian Higgs-Kibble model,  Comm. Math. Phys., {\bf 42}, 127 (1975).


\bibitem {Tyutin} I.V. Tyutin, Gauge invariance in field theory and statistical physics in operator formalism,
Lebedev Institute preprint N39 (1975), arXiv:0812.0580.


\bibitem{Landau} L.D. Landau and E.M. Lifshitz, {\it The Classical Theory of Fields,} 4'th ed., Pergamon Press, New York, 1975.

\bibitem{Weinberg1} S. Weinberg, {\it Gravitation and Cosmology: Principles and Applications of the General Theory of 
Relativity,} John Wiley \& Sons, New York, 1972.

\bibitem{Noether} E. Noether, Invariante Variationsprobleme, Nachr. Ges. Wiss. G\"ott., Math.-Phys.,  Klasse 235 (1918),
arXiv:physics/0503066.


\bibitem{Belinfante} F.J. Belinfante, On the spin angular momentum of mesons, Physica {\bf 6}, 887 (1939):
On the current and the density of the electric charge, the energy, the linear momentum and the
angular momentum of arbitrary fields, Physica {\bf 7}, 449 (1940).  

\bibitem{Rosenfeld} L. Rosenfeld, Sur le tenseur d'impulsion-\'energie, M\'emoires Acad. Roy. Belg. Sci., {\bf 18}, 1 (1940).


\bibitem{Weinberg2} S. Weinberg, {\it The Quantum theory of fields} Vol. I, Cambridge Univ. Press, Cambridge, UK
(1995); p. 316. 

\bibitem{Munoz} G. Mu\~noz, Lagrangian field theories and energy-momentum tensors, Am. Jour. Phys. {\bf 64}, 1153 (1996).

\bibitem{Saravi} R.E. Gamboa Saravi, On the energy-momentum tensor, Jour. Phys. A {\bf 37}, 9573 (2004).

\bibitem{Pons} J.M. Pons, Noether symmetries, energy-momentum tensors, and conformal invariance in classical field
theory, J. Math. Phys. {\bf 52}, 012904 (2011).

\bibitem{Blaschke} D.N. Blaschke, F. Gieres, M. Reboud, and M. Schweda, The energy-momentum tensor(s)  in classical
gauge theories,  Nucl. Phys.  B, {\bf 912}, 192 (2016).

\bibitem{Ilin}  R.V. Ilin and S.A. Paston, Exact relation between canonical and metric energy-momentum tensors
for higher derivative tensor field theories, Eur. Phys. J. Plus {\bf 134}, 21 (2019).

\bibitem{Freese} A. Freese, Noether's theorems and the energy-momentum tensor in quantum gauge theories,
Phys. Rev. D, {\bf 106}, 125012 (2022).

\bibitem{Gotay} M.J. Gotay and J.E. Marsden, Stress-energy-momentum tensors and the Belinfante-Rosenfeld formula,
Contemp. Math. {\bf 132}, 367 (1992).

\bibitem{Forger} M. Forger and H. R\"omer, Currents and the energy-momentum tensor in classical field theory:
a fresh look at an old problem, Ann. Phys. {\bf 309}, 306 (2004).

\bibitem{Baker} M.R. Baker, N. Kiriushcheva, and S. Kuzmin, Noether and Hilbert (metric) energy-momentum tensors are not, in general, equivalent,
Nucl. Phys.  {\bf B}, 962, 115240 (2021).

\bibitem{Sanchez} M. S\'anchez, On the geometry of static spacetimes, Nonlinear Analysis, {\bf 63}, e455 (2005).

\bibitem{Fulling} S.A. Fulling, Nonuniqueness of canonical field quantization in Riemannian spacetime, Phys. Rev. D, {\bf 7}, 2850 (1973).

\bibitem{Derezinski} J. Derezi\'nski and D. Siemssen, Feynman propagators on static spacetimes, Rev. Math. Phys.  {\bf 30}, 1850006     (2018).

\bibitem{AW1} H.A. Weldon, Fermions without vierbeins in curved space-time, Phys. Rev. D, {\bf 63}, 104010 (2001).


\end{thebibliography}
\end{document}